\documentstyle[prl,aps,twocolumn,epsf,floats,psfig]{revtex}

\draft

\begin{document}

\twocolumn[\hsize\textwidth\columnwidth\hsize\csname
@twocolumnfalse\endcsname

\title{A 4-unit-cell superstructure in optimally doped
YBa$_2$Cu$_3$O$_{6.92}$ superconductor}

\author
{Zahirul Islam$^{1,}$\cite{ZI} \and X. Liu$^2$ \and S. K. Sinha$^2$
\and J. C. Lang$^1$ \and S. C. Moss$^3$ \and D. Haskel$^1$
\and G. Srajer$^1$ \and P. Wochner$^4$ \and D. R. Lee$^1$
\and D. R. Haeffner$^1$ \and U. Welp$^5$}
\address
{$^1$Advanced Photon Source, Argonne National Laboratory,
Argonne, IL 60439}
\address
{$^2$Department of Physics, University of California, San Diego,
CA 92093}
\address
{$^3$Department of Physics and Texas Center for Superconductivity
and Advanced Materials, University of Houston,
TX 77204}
\address
{$^4$Max-Planck-Institut f\"ur Metallforschung, 70569 Stuttgart,
Germany}
\address
{$^5$Materials Science Division, Argonne National Laboratory,
Argonne, IL 60439}

\date{\today}

\widetext

\maketitle 

\begin{abstract}
Using high-energy x-ray diffraction we show that a 4-unit-cell
superstructure, {\bf q}$_0$=$\left(\frac{1}{4},0,0\right)$,
along the shorter Cu-Cu bonds coexists with superconductivity
in optimally doped YBCO. A complex set of anisotropic atomic
displacements on neighboring CuO chain planes, BaO planes, and CuO$_2$
planes, respectively, correlated over $\sim$3-6 unit cells
gives rise to diffuse superlattice peaks. Our observations
are consistent with the presence of Ortho-IV nanodomains
containing these displacements.
\end{abstract}


\vskip0.5pc]
\narrowtext

There have been several experimental indications recently of 
inhomogeneous phases of various kinds developing in the high-$T_c$
cuprates. These were predicted on purely electronic
considerations\cite{stripes},
and were given credibility with the discovery, by neutron
scattering, of the so-called ``striped phase'' in the low-temperature
tetragonal La$_{1.6-x}$Nd$_{0.4}$Sr$_x$CuO$_4$\cite{jmt}.
In this phase the magnetic periodicity is equal to twice that of
the charge. Later neutron measurements\cite{lake01} have 
revealed incommensurate spin density waves with a period close
to 8 unit cells along the {\bf a} axis at optimal doping in the
superconducting phases of La$_{2-x}$Sr$_x$CuO$_4$ (LSCO) associated
with the Cu spins in the CuO$_2$ planes, which are enhanced inside
the vortices induced by an applied magnetic field\cite{lake01}.
Recent STM measurements\cite{stm-exp} in Bi-2212 reveal a spatial
modulation of the local electronic density of states possibly
arising from quasiparticle interference between states
that are nested across the Fermi surface. Very recent
Josephson tunneling work suggests a non-uniform
superconducting condensate in LSCO\cite{Basov}.
Inelastic neutron scattering
has also revealed the existence of phonon anomalies at a wavevector
{\bf q}$_0$=($\frac14$,0,0), in reciprocal lattice units,
$\frac{2\pi}{a}$, in both YBa$_2$Cu$_3$O$_{6+x}$ (YBCO) and LSCO
compounds\cite{ybco-eph,e-ph}. The role of lattice strains in
creating texture has also been discussed
recently\cite{Zhu03}. In addition, in the YBCO
family of compounds, there are O vacancies in the
CuO chains (except at stoichiometry, {\it i.e.} x=1.0, sligthly
beyond the optimum doping), which order to some degree
in well-prepared samples\cite{oxyord}. The phase diagram of the
vacancy ordering on the CuO chains has been theoretically
discussed by De Fontaine and coworkers\cite{asynnni}.
While the exact origin of inhomogeneities is being debated,
diffraction studies to understand the nature of such phases
are absolutely necessary.

\begin{figure}[h!t]\centering
\epsfxsize=70mm
\epsfbox{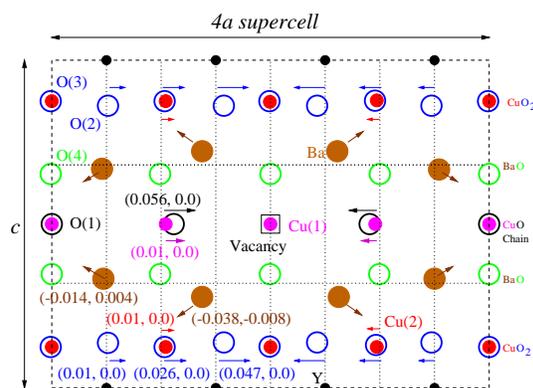}
\caption{An ideal atomic displacement (arrows) pattern at $\sim$7 K.
All the atoms have been projected on the {\it ac} plane. Note that
primary diplacements ($\delta{\bf u}$'s) are along the {\bf a} axis,
{\it i.e.} along the shorter Cu-Cu bond direction. $\delta{\bf u}$'s,
$(\delta_a,\delta_c)$ in units of $a$ and $c$,
of respective atoms are given in parentheses. $\delta{\bf u}$'s of all
other atoms are related by mirror symmetry.}
\label{cell-model}
\end{figure}

We have used high-energy synchrotron x-ray scattering to look for the 
occurrence of inhomogeneous phases and lattice modulations
in the YBCO family of compounds. In an underdoped
YBCO (x$\approx$0.63) compound we previously found modulations with
{\bf q}$_0$ = ($\sim\frac25$,0,0)
possessing short-range order\cite{ZI02} which
coincided with a harmonic of the so-called Ortho-V
phase of O-vacancy ordering on the Cu-O chains\cite{asynnni} with a
5-unit-cell repeat along the {\bf a} axis.
However, the intensities of the diffuse satellites
clearly showed that displacements of atoms in the CuO chain planes,
the CuO$_2$ planes and the BaO planes were involved\cite{ZI02}. This
paper describes similar measurements carried out on an optimally 
doped YBCO crystal, possessing on the average 8\% of O 
vacancies on the chains. We find that a 4-unit cell
(Ortho-IV) superstructure, {\bf q}$_0$=($\frac14$, 0, 0),
involving correlated displacements of atoms (Fig.\ \ref{cell-model}),
indeed coexists with superconductivity.

For this study, a high-quality {\it detwinned} crystal
($\sim$1 mm$\times$1 mm$\times$130 $\mu$m) of optimally
doped YBCO ($T_c=91.5$ K, $\Delta T_c\approx 1$ K) was chosen. The use
of a detwinned crystal allowed us to determine unambiguously
the anisotropy of scattering,
the direction of modulation, and the polarization of atomic
displacements, respectively. The crystal was annealed
at 420$^\circ$C in flowing pure O$_2$ for about a week and was stress
detwinned in flowing O$_2$ at the same temperature.
Polarization-sensitive optical microscopy showed the presence of a
single twin domain. The crystal mosaic was
$\sim0.03^\circ$. The {\bf c} axis was
perpendicular to the large crystal facet. High-energy (36 keV) x-ray
diffraction studies were performed on the 4ID-D beamline at the
Advanced Photon Source. Experimental details
can be found elsewhere\cite{ZI02}.

\begin{figure}[h!t]\centering
\epsfxsize=75mm
\epsfbox{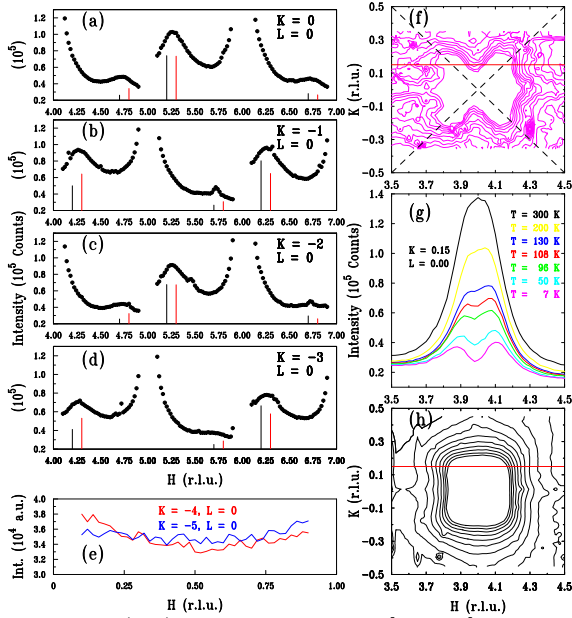}
\caption{
(a-d) Raw data showing [H, 0, 0] scans for
several integer values of K and L=0. Satellite peaks
correspond to {\bf q}$_0$ = ($\frac14$, 0 ,0). Vertical
lines (displaced along H for clarity) compare the observed (black)
and calculated (red) intensities corrected for
geometric factors; (e) [H, 0, 0] scans with high 
values (odd/even) of K relative to H showing the absence of
{\bf q}$_0$ satellites above background; (f) Contour plot of diffuse
intensity at 7 K around (4, 0, 0) Bragg peak; diagonal dashed lines
indicate [$\pm1$,$\pm1$,0] directions; (g) Line scans at
different $T$ showing how TDS overwhelms HDS above $\sim$200 K;
(h) Contour of diffuse intensity at 300 K.
Red line in (f) and (h) indicates where the linescans displayed
in (g) were taken.
}
\label{hscans}
\end{figure}

\begin{figure}[h!t]\centering
\epsfxsize=75mm
\epsfbox{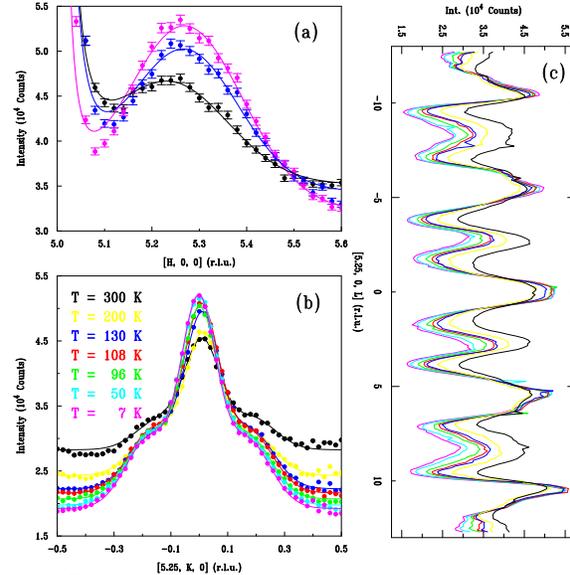}
\caption{
(a) [H, 0, 0] scans at several temperatures.
Lines are fits using a combination of a  Lorentzian (TDS), a Gaussian
(satellite), and a constant term. (b) K scans through
the (5.25, 0, 0) peak. (c) Intensity modulations along {\bf c*} of the
same peak. Note that the oscillation amplitude grows on decreasing $T$.
Different $T$'s are shown with unique colors (b).
}
\label{huang}
\end{figure}

Figs.\ \ref{hscans}(a)-(d) show several {\bf a}-axis [H, 0, 0] scans
normal to the Cu-O-Cu chain direction for different integer values
of $K$ taken at $\sim$7 K. Broad satellite peaks corresponding to
{\bf q}$_0 = \left(\frac14, 0, 0\right)$ are clearly visible near
Bragg peaks, $(h,k,0)$, when $h$ and $k$ have mixed parity.
When $h$ and $k$ have the same parity
strong thermal diffuse scattering (TDS), however, overwhelms the
superlattice peaks. The intensity of the peak at
{\bf Q}={\bf G}+{\bf q}$_0$ ({\bf G} is a reciprocal lattice vector) is
$\sim$450 photons/second which is some $\sim$10$^6$
orders-of-magnitude weaker than that of a Bragg peak. The width in the
{\bf a}* direction of {\bf q}$_0$ satellites is much larger than the
resolution, indicating a very short correlation range
($\xi_a\sim3a$, using Scherrer formula\cite{Warren})
along the {\bf a} axis. The use of a detwinned crystal made it
unambiguous that the modulation vector is
{\bf q}$_0$=$\left(\frac14,0,0\right)$ and {\it not}
$\left(0,\frac14,0\right)$. The width in the {\bf b}-axis
direction (Fig\ \ref{huang}(b)) corresponds to larger correlation
length $(\xi_b\sim6b)$, while modulations of the intensity which
extend along the ${\bf c^*}$ axis through the satellite peak as
shown in Fig.\ \ref{huang}(c) indicate correlations only between
neighboring Cu-O chain planes, BaO planes, and CuO$_2$
planes (Fig.\ \ref{cell-model}), respectively, as  obtained from
Fourier transform (Patterson function) of these intensity patterns
(this is very similar to the results found in the underdoped
system\cite{ZI02}).

In addition, significant lattice-strain effects are present in this
material. A 2-dimensional scan around (4, 0, 0), as shown in
Fig.\ \ref{hscans}(f), reveals a strongly anisotropic ``bowtie''-shape
Huang diffuse scattering (HDS) pattern with lobes extending along the
four [$\pm$1, $\pm$1, 0] directions. The two superlattice
peaks at (4$\pm\frac14$, 0, 0) can be discerned, although
they are not completely resolved. A set of linescans through the
diffuse lobes at several temperatures is shown in
Fig. \ref{hscans}(g). Whereas at low temperature two broad peaks
corresponding to two lobes are clearly visible, on increasing $T$
the peaks become indiscernible from the rapidly growing TDS above
$\sim$200 K. Room-temperature pattern of diffuse scattering
(Fig.\ \ref{hscans}(h))
is nearly identical to TDS around (4, 0, 0) calculated using elastic
constants of YBCO (not shown). Earlier x-ray studies\cite{jian91} of
tetragonal system YBa$_2$(Cu$_{0.955}$Al$_{0.045}$)$_3$O$_7$
showed that HDS arises from shear distortions
due to long-wave fluctuations of O concentrations in the
chains along {\bf a} and {\bf b} axes. It is possible that the O
stoichiometry in the CuO chains is non-uniform on submicron scales
due to the formation of defective short-range O-ordered
domains discussed below.

A 4-unit-cell periodic
(Ortho-IV) phase is expected near O stoichiometry of
6.75\cite{asynnni} ({\it i.e.} one out of every four CuO
chains has no O atoms denoted by $\langle1101\rangle$)
whereas in optimally doped material the
stoichiometry is 6.92 ({\it i.e.} approximately one out
of every ten CuO chains has vacancies). There are two
ways to explain the formation of $\langle1101\rangle$ structure
near optimum doping. If O concentration is nonuniform within the
CuO-chain planes then vacancies tend to phase separate within
the formation range of the Ortho-IV phase\cite{asynnni}. Secondly,
if the long-range Coulomb interactions among distant-neighbor
vacancies are not negligible then the $\langle1101\rangle$ phase can
be stable even near the optimal doping with dilute concentration of
vacancies. In both cases, however, the ordering will be short ranged
and imperfect, leading to significant lattice strains responsible
for HDS.

Next, we note some general features of our data which were used to
narrow down possible models of atomic displacements
($\delta{\bf u}$'s). First, a strong intensity
asymmetry between the +{\bf q}$_0$ and -{\bf q}$_0$ satellites
is observed around all Bragg points. A strong asymmetry can occur
if $\delta{\bf u}$'s are large\cite{RWJ48},
or as a result of destructive interference
between diffuse scattering due to disorder and displacive
modulation\cite{guin63} as found in quasi-1D charge density wave
systems\cite{braz97,rouz00}. Secondly, for a given satellite
at ($h$,$k$,0)+{\bf q}$_0$ the intensity is either very strong or weak
when $h$ and $k$ have the same or mixed parity,
respectively. This implies out-of-phase
displacements of the dominant scatterers. Thirdly, no second harmonic
(2{\bf q}$_0$) satellites were observed indicating essentially a
sinusoidal modulation. Finally, scans
(Fig.\ \ref{hscans}(e)) such as [H, -5, 0] (H $\in [0.1-0.9]$) found no
superlattice peaks suggesting the absence of any
$\delta{\bf u}\parallel{\bf b}$ associated with the
{\bf q}$_0$ modulation. We performed calculations
without assuming any displacements to be small in the presence of an
Ortho-IV phase in the CuO-chain plane.

The intensity calculations were performed using
\begin{eqnarray*}
I_{\text{diff}}({\bf Q})\propto\left|\sum_{n}
f_n(Q) e^{-W_n(Q)}
e^{-i{\bf Q}\cdot({\bf R}_n+\delta{\bf u}_n)}\right|^2
\end{eqnarray*}
where the displacement relative to an  average lattice
site (${\bf R}_n$) of the $n$-th atom is $\delta{\bf u}_n$,
\noindent $f_n(Q)$ and $e^{-W_n(Q)}$ are the form factors and
Debye-Waller factors (DWFs), respectively. The summation was carried
out only over Cu, O, and Ba atoms in the $4\times1\times1$
supercell (Fig.\ \ref{cell-model}).
The expression above is for integrated intensity of the satellites
regardless of peak widths. The extraction of integrated intensity from
the experimental data at $\sim$7 K, however, was difficult due to
the presence of TDS and HDS.
Nevertheless, since the satellite peaks are sharper
than TDS and HDS, and located away from Bragg
peaks, it is possible to model the satellites at the lowest $T$ using
a Gaussian above some monotonic background. We found that extracted
intensities for the strong peaks varied $\sim$15-20\% ($\sim$35\% for
the weak peaks) depending on how the background scattering was modeled.
A least-squares procedure was performed taking these errors and
the general considerations discussed above into account
to fit the intensities of 45 peaks within the [H, K, 0] zone
and the intensity modulation of the (5.25, 0, 0) peak along {\bf c*}.
We kept the displacements symmetric about the
Cu(1)-O(4)-Cu(2) and CuO mirror planes centered on the O-vacancy
inside a supercell. Vertical bars in Figs.\ \ref{hscans}(a)-(d)
indicate that there is good agreement between the calculated (red)
and observed (black) intensities within experimental uncertainties.
The model obtained from fitting is shown in Fig.\ \ref{cell-model}.
While the dominant contributions come from displacements of Ba
and Cu atoms, both chain (O(1)) and plane oxygen atoms (O(2) and O(3))
are displaced significantly. Note that the displacements are primarily
along the {\bf a} axis. Although there may be small
displacements along the {\bf c} axis as well, we are more
certain of them in the case of Ba.
Our error estimates are $\sim10-15$\% for Ba and Cu
$\delta{\bf u}$'s, and $\sim15-25$\% for O atoms, respectively.
While the model obtained may not be perfect given the difficulty
of extracting accurate intensities, it does account for
all the systematics of the data. Furthermore, it portrays
a pattern of displacements similar to that of Ortho-V phase in
an underdoped YBCO obtained from first-principles electronic
calculations\cite{DdF03}. In our case, however, the periodicity
is 4$a$ (Fig.\ \ref{cell-model}) along the {\bf a} axis.

\begin{figure}[h!t]\centering
\epsfxsize=75mm
\epsfbox{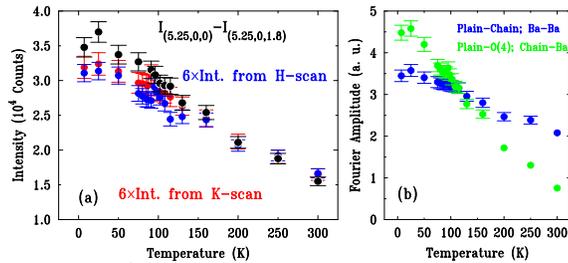}
\caption{
(a) Temperature dependence of (5.25, 0, 0) peak.
(b) $T$ dependence of Fourier amplitudes
obtained from intensity modulations shown
in Fig.\ \ref{huang}(c).}
\label{dfpk-t}
\end{figure}

Fig.\ \ref{huang}(a) shows [H, 0, 0] scans
through a superlattice peak at several temperatures. It is clear from
these scans that as $T$ is increased the intensity of the {\bf q}$_0$
peak decreases relative to the TDS emanating from (5,0,0).
In fact, an inspection of the data
reveals that the intensity (area under the dome) at 7 K is at
least twofold larger than that at 300 K. Intensity
modulation of (5.25, 0, 0) peak presented in Fig.\ \ref{huang}(c)
shows that while the mean intensity of the oscillations falls with
decreasing $T$ due to the reduction of TDS, the oscillation amplitude
about the mean grows. In order to get more quantitative information
as a function of temperature we fitted\cite{ZI02} a combination of
a Lorentzian (TDS), a Gaussian, and a constant term to the H-scans
(Fig.\ \ref{huang}(a)). K-scans shown in Fig.\ \ref{huang}(b)
are well represented by a combination of three Gaussian line profiles,
one for the central satellite peak and two for the
broad lobes, respectively, and a constant term to
account for the background which is predominantly made up of TDS.
Note that both HDS and TDS contribute to the
broad lobes. Since the peak widths and positions do not
change with $T$ only the peak heights and the constant term
({\it i.e.} four parameters all together) were needed to
fit the entire data. Fig.\ \ref{dfpk-t}(a) shows the $T$-dependence
of the fitted intensity for (5.25, 0, 0) peak. Although keeping
positions and widths constant may introduce some systematic errors
for the central satellite, its integrated intensity
(width$\times$peak intensity) agrees well with that obtained
in fitting the H-scan as shown in Fig.\ \ref{dfpk-t}(a).
The intensity was also estimated via the
maximum amplitude of the modulation defined as
$I_{(5.25,0,0)}-I_{(5.25,0,1.8)}$. All three cases consistently
show that the superlattice peak decreases nearly
linearly with increasing $T$ (Fig.\ \ref{dfpk-t}(a)).
If this linear trend continues then the
intensity will extrapolate to zero around $\sim$500 K.
Furthermore, Fig.\ \ref{dfpk-t}(b) shows Fourier amplitudes obtained
from intensity modulations (see Fig.\ \ref{huang}(c)), which are a
measure of displacement-displacement correlations as a function of $T$.
It is clear that both amplitudes also grow stronger at lower
temperatures.

Although the origin of {\bf q}$_0$ can be attributed to Ortho-IV phase
it is puzzling to observe a large increase of the diffuse peak with
decreasing $T$. Using the displacement model presented above and
DWFs for the average lattice measured on ceramic
samples \cite{DWF} we estimated
$\frac{I(7 \text{K})}{I(300 \text{K})}\approx1.2$ for the intensity
of (5.25, 0, 0) satellite, which is at odds with the observed ratio
of at least $\sim$2.2. Since diffusive motion of chain oxygens (O(1))
practically freezes below $\sim$ 250 K,
the growth of Ortho-IV patches in size or number seems unlikely.
Given that atomic displacements (Fig.\ \ref{cell-model})
are clearly anharmonic in these nanoscale patches it
appears that enhanced elastic softening of the lattice takes
place within these regions on lowering $T$ which
may account for the low-$T$ increase of the intensity.

In conclusion, we have shown that lattice modulations
with a 4-unit-cell periodicity exist from above room temperature
down to the lowest temperatures in optimally doped YBCO.
These correspond to local regions in extent $\sim$3-6 unit cells in
the $ab$ plane and less than one unit cell along the {\bf c} axis.
From the $\delta{\bf u}$'s (Fig.\ \ref{cell-model}) one may
calculate DWFs for the whole crystal and by comparison with the
experimental DWFs\cite{DWF} we
estimate roughly $\sim$10-20\% of the crystal
contain these patches at the lowest $T$.
At low temperatures clear evidence of anisotropic
strain in the lattice is provided by anisotropic 
patterns of  HDS around the Bragg points. This HDS originates
with the strain induced by the disorder between O atoms and vacancies
along both {\bf a} and {\bf b} axes; strains induced by the patches
is not assumed. However, the coincidence of the observed periodicity
({\bf q}$_0$=($\frac14$,0,0)) with values expected for charge
instabilities in the CuO$_2$ planes from measurements on other
superconducting cuprates is striking.
It seems clear that in YBCO the electronic structure and the 
oxygen vacancies together possess instabilities which can lead to  
inhomogeneous phases and local softening of the lattice.

We have benefitted from our discussions with B. W. Veal, D. de
Fontaine, V. Ozolins and D. Basov. Use of the Advanced Photon Source is
supported by the U.S. Department of Energy, Office of Science,
Office of Basic Energy Sciences, under Contract No. W-31-109-ENG-38.
SCM thanks the NSF for support on DMR-0099573.

\end{document}